\documentstyle[aps,twocolumn]{revtex}

\begin{document}
\draft
\title{Motion tomography of a single trapped ion}
\author{J. F. Poyatos$^\dagger$, R. Walser, J. I. Cirac$^\dagger$, P. Zoller}
\address{Institut f{\"u}r Theoretische Physik,
Universit{\"a}t Innsbruck, 6020
Innsbruck, Austria}
\author{R. Blatt}
\address{Institut f\"ur Experimental Physik, 
Universit{\"a}t Innsbruck, 6020
Innsbruck, Austria}
\date{December 22, 1995}
\maketitle

\begin{abstract}
A method for the experimental reconstruction of the quantum state of
motion for a single trapped ion is proposed. It is based on the measurement
of the ground state population of the trap after a sudden change of the trapping
potential. In particular, we show how the $Q(\alpha)$ function and the
quadrature distribution $P(x,\theta)$ can be measured directly. In an example
we demonstrate the principle and analyze
the sensibility of the reconstruction process to experimental uncertainties 
as well as to finite grid limitations. Our method is not restricted to the 
Lamb--Dicke Limit and works in one or more dimensions.
\end{abstract}


\narrowtext


The central entity of quantum physics is the density operator $\rho$.
It contains all measurable information about the state of a system that
can be obtained according to the principles of quantum physics.
Recent theoretical advances established constructive procedures to recover 
the full information about the state of a system from the repeated 
measurement of a complete set of observables. 
From the experimentally detected probabilities 
\begin{mathletters}
\begin{eqnarray}
\label{Qfun} 
Q(\alpha)&=&\frac{1}{\pi}\mbox{Tr}[\rho\, |\alpha\rangle \langle\alpha|]
=\frac {1}{\pi} \langle \alpha| \rho | \alpha \rangle,  \\
\label{pxteta}
P(x,\theta)&=&\mbox{Tr}[\rho\, |x,\theta\rangle \langle x,\theta|]
=\langle x, \theta |\rho |x,\theta \rangle, 
\end{eqnarray}
\end{mathletters}
one can determine the state $\rho$ uniquely. Here $|\alpha\rangle$ 
denotes a coherent state and $|x,\theta\rangle$ is a quadrature 
eigenstate. 
So far, the underlying theory has been developed for finite dimensional 
discrete systems \cite{New68,Leonhardt95}, 
like spin- or angular momentum states, as well as
for continuous systems. 
This approach is generally referred to as phase space tomography \cite{Vo89}.

One of the most important and beautiful applications has been the
tomographic measurement of the Wigner function for a single mode of the
electro-magnetic radiation field by Raymer {\em et.al.} \cite{Sm93} 
following the proposal by Vogel and Risken \cite{Vo89}.  It is based on 
a measurement of the quadrature distribution $P(x,\theta)$ with the 
help of a homodyne technique. On the other hand, the $Q$ function 
has been measured recently in various experimental schemes, using 
well--known techniques of photo detection, together with related
experiments of phase measurement \cite{Le95}.

Apart from cavity QED, a single trapped ion is the other 
testing ground of 
the intriguing features of quantum mechanics \cite{Trapreview}.  
The motion of a single trapped ion can
be easily modified using laser light, and decoherence in such a system
can be made nearly negligible during long times.  Using these
properties, several proposals have emerged dealing with the
preparation of non--classical states of motion.
Just recently, the first
observation of non--classical states such as Fock states, and squeezed
states has been reported \cite{NISTExperim}.  
Hence,
the next step of research is to characterize these
states.  Given the analogy between cavity QED and a
trapped ion interacting with a laser, one could imagine that some
techniques developed in the framework of cavity QED can be immediately
transcribed to the ion system. For example, one can characterize the
motional state by measuring the evolution of the ion population inversion 
\cite{Bl92,Ci93a,Schendos,toschek95}. Endoscopic techniques, for example,
permit a complete state detection if there is no statistical uncertainty
in the state preparation process. Unfortunately,
this method does not allow to recover the whole density matrix describing
the ion motion. Moreover, the mentioned analogy is only valid
in the Lamb--Dicke regime, whereby the motion of the ion is restricted
to a region smaller than a wavelength, which limits the applicability of these
methods. Thus, it would be desirable to have a method to recover 
the full information about the motional state of an ion 
valid for more general situations.

In the present letter we propose a novel realization of a phase space 
tomography to determine the motional state of an ion in a harmonic 
trapping potential. In contrast to a recent proposal \cite{Vo95} that
addressed the same question, our scheme is not restricted to the 
Lamb-Dicke regime and can be extended easily  to more than one spatial
dimension. Furthermore,  an implementation of this idea is feasible 
with present experimental setups. 
Specifically, we will present procedures:
\begin{description}
\item[(i)] to measure the $Q(\alpha)$ function, and
\item[(ii)] the quadrature distribution function $P(x,\theta)$.
\end{description}

Our model consists of a single ion, trapped in a harmonic potential oscillating with a frequency $\nu$.  
The internal structure of the ion will be specified later in the context of the measurement of the motional state. 
We use a density operator $\rho$ to describe this unknown  state of the 
particle and represent it in the Fock basis of the harmonic oscillator, i.e.
\begin{equation}  
\label{density}
\rho = \sum_{n,m=0}^\infty \rho_{nm} |n\rangle \langle m| .
\end{equation}

Let us first show how the $Q$ function given in Eq.~(\ref{Qfun}) can be measured
experimentally. For this purpose, we re-express the $Q$ function as
\begin{eqnarray}  
\label{Qfuntwo}
Q(\alpha)&=&\frac {1}{\pi} \langle 0|\tilde \rho |0 \rangle,\\
%
\mbox{where }\quad \quad \quad \tilde \rho&=&U(|\alpha|,\theta)\, \rho\, U^\dagger(|\alpha|,\theta).
%
\end{eqnarray}
Here, $U^\dagger(|\alpha|,\theta)={\cal R}^ \dagger (\theta) 
{\cal D}(|\alpha|)$ is the unitary transformation that is given by the 
displacement operator ${\cal D} (\alpha)=\exp(\alpha a^\dagger - 
\alpha^* a)$ and the phase shifting operator ${\cal R}(\theta)=
\exp (-i \theta a^\dagger a)$ that acting on the vacuum create a 
coherent state $\left||\alpha|\,e^{i\theta}\right>=U^\dagger 
(|\alpha|,\theta)\left|0\right>$. As usual, $a^\dagger$ and 
$a$ denote creation and annihilation operators that obey  $[a,a^\dagger]=1$.

According to Eq.~(\ref{Qfuntwo}), one has to determine the probability 
of the state represented by $\tilde \rho$ to be in the ground state 
of the harmonic potential, in order to measure this $Q$ function.
Note that such a state is related to the original $\rho$ by the 
unitary transformation $U(|\alpha|,\theta)$. Consequently, the
identification of this transformation with a physical process would 
enable us to measure the $Q$ function.  In the context of an ion 
trapped in a harmonic potential, this identification is as follows.  
The operator ${\cal R}(\theta)$ corresponds to the free evolution, 
whereas the operator ${\cal D} (\alpha)$ corresponds to a sudden 
displacement of the harmonic trap.  A pictorial representation of 
these operations in phase space is shown in Fig.~1(a). Thus, in order to
measure the $Q$ function in a trap, one simply has to perform the following
steps:  
{\bf(i)}~Wait a particular time $t$ while the ion evolves freely in the
trap. This gives it the appropriate phase shift according to $\theta=\nu t$.
{\bf (ii)}~Suddenly displace the center of the trap to the right for a distance $d$, so
that $| \alpha |= \sqrt{\frac{m \nu}{2 \hbar} } d $.
{\bf (iii)}~Finally, measure the probability of the ion to be in the lowest motional state $|0\rangle$.  

To achieve this last
step, one may use the internal structure of the ion.  Typically, it consists of
three levels $|g\rangle$, $|e\rangle$, and $|r\rangle$, where $|g\rangle
\rightarrow |r\rangle$ is a dipole--forbidden transition or Raman transition,
whereas $|g\rangle \rightarrow |e\rangle$ is a dipole--allowed transition.
Initially, the ion is in the internal ground state $|g\rangle$.  After the step
{\bf (ii)} a laser beam is tuned to the lower side--band of the $|g\rangle \rightarrow |r\rangle$ internal
transition.  One can then transfer completely the population of the ground
states $|n,g\rangle$ (with $n=1,2,\ldots$) to the excited states $|n-1,r\rangle$
coherently as described in Ref.~\cite{Ci94a} by an adiabatic sweep of the laser
frequency.  After this population transfer one can switch on another laser, this
time on resonance with the transition $|g\rangle \rightarrow |e\rangle$ [see
Fig.~1(c)]. The appeareance of fluorescence indicates the presence of 
population  in the $|0,g\rangle$ state.
One can repeat the same sequence of steps in
order to determine the probability of the ion to be in the ground state.  An
alternative (more sophisticated) way of measuring this probability may be
achieved by detecting the collapses and revivals in the population inversion,
since this technique provides the whole population of the Fock states
\cite{Bl92}.

Up to now, we have not addressed the question of the final reconstruction of our state from the
experimental data of the $Q$ function that is obtained in this manner.  In principle, one
could use the method \cite{Gardiner} 
that relates this function to the matrix elements
$\rho_{n,m}$.  This is, however, impractical, since it requires 
the $n$--th and $m$--th derivatives of the $Q$ function, i.e.  their knowledge 
over a continuous interval.  Another possibility would be to assume that
$\rho_{n,m}=0$ for $n,m>n_{\rm max}$, for a given $n_{\rm max}$. 
The measurement of $Q(\alpha_i)$
for $n_{\rm max}^2$ (independent) values of $\alpha_i$ would allow us to find $\rho_{n,m}$
by simple matrix inversion.  This procedure is also
of limited usefulness since small 
deviations from the exact values of 
the $Q$ function (such as experimental uncertainties) 
give large errors in the
reconstruction.  This is due to the fact that $Q$ is the smoothest function of
all $s$--parameterized quasi-distributions.

An alternative way of reconstructing the state of a quantum system is
by means of quantum tomography. Tomography is an  experimental tool used in
several areas of research which allow us to reconstruct an unknown object
from measured data. In the context of quantum physics, the data we will have to measure
are the so--called quadrature distribution functions given by Eq.~(\ref{pxteta})
where $|x,\theta\rangle={\cal R}^\dagger(\theta)|x\rangle$ is the eigenstate of the operator 
${\hat x}(\theta)={\cal R}^\dagger (\theta){\hat x}{\cal R} (\theta)$, with eigenvalue $x$ 
(${\hat x}$ is the dimensionless position operator of the harmonic oscillator
$m\nu/\hbar\rightarrow1$).  
This distribution is equivalent to that given by the
marginal distribution for quadrature components using the Wigner function
description of the state \cite{Le95}.  

Our scheme for the measurement of $P(x,\theta)$ is based on the well--known property
of the squeezed states
\begin{equation}
\label{line}
|x,\theta \rangle =\lim_{|\epsilon |\rightarrow \infty }
{\cal N}_{\epsilon}
|\alpha ,\epsilon\rangle,
\end{equation}
where $|\alpha ,\epsilon \rangle ={\cal D}(\alpha){\cal S}(\epsilon )|0\rangle
$, ${\cal S}(\epsilon )=\exp[(\epsilon ^{*}a^2-\epsilon a^{\dagger ^2})/2]$ is
the ``squeeze'' operator, $\epsilon =|\epsilon |e^{2i\theta }$ and
$\alpha=x e^{i\theta}/\sqrt{2}$ ($0\le\theta<\pi$).  
As proper position eigenstates are not normalizeable, there is a
constant of proportionality ${\cal N}_{\epsilon}=(\exp{(2
|\epsilon|)}/(4\pi))^{\frac{1}{4}}$ that increases with the degree of
squeezing. As before, we can use these states to reexpress the quadrature
distribution in the form
%
%
\begin{eqnarray}
P(x,\theta )&=&\lim_{|\epsilon |\rightarrow \infty }
|{\cal N}_{\epsilon}|^2
\langle 0 | {\tilde \rho} | 0\rangle.\\ 
\mbox{Here }\quad \quad \quad  \tilde \rho& =& 
U(|\epsilon|,|\alpha|,\theta)\, \rho \,U^\dagger(|\epsilon|,|\alpha|,\theta),
\end{eqnarray}
where 
$U^\dagger(|\epsilon|,|\alpha|,\theta)=
{\cal R}^\dagger(\theta){\cal D}(|\alpha|) {\cal S}(|\epsilon|)$ denotes the operation that creates a squeezed state and we used furthermore the
property
${\cal S}(\epsilon )$= ${\cal R}^{\dagger }(\theta ){\cal S}(|\epsilon |){\cal R}(\theta )$.  
Thus, to measure the quadrature distribution one
has to find the physical processes that correspond to the unitary operators
${\cal R}$, ${\cal D}$ and ${\cal S}$.  The first two are the same as those needed for
the experimental determination of the $Q$ function.  On the other hand, it is
well known that sudden changes in the frequency of a harmonic oscillator lead
to squeezed states, a process that can be readily achieved in a
trap, just opening or closing the harmonic potential \cite{Ja86}.  In
particular, changing the trap frequency from $\nu$ to $\nu'$ leads to a squeezing
parameter $|\epsilon |$=$\frac 12 \mbox{ln}{\frac \nu {\nu^ \prime}}$.  

Thus, in order to measure the quadrature distribution in our trap
one has to follow these four steps:  
{\bf (i)}~Wait for a time $t$, such as $\theta=\nu t$.
{\bf (ii)}~Perform a sudden displacement of the center of the trap to
the right a distance $d$, so that $| \alpha |= \sqrt{\frac{m \nu}{2 \hbar} } d $.
{\bf (iii)}~Change the trap frequency instantaneously from $\nu$ to $\nu'$.
{\bf (iv)}~Determine the population of the motional ground state.
Note that the steps {\bf (i,ii)} and {\bf (iv)} are the same as before.

We are now in position to extract the full information about the unknown quantum
state starting from the quadrature distribution.  This can be done as in the
case where one measures the quantum state of light by means of balanced homodyne
detection. As has been shown by Vogel an Risken \cite{Vo89}, one
can reconstruct the Wigner function $W(x,p)$ by means of the so--called inverse Radon
transformation. Alternatively, one can use one of the algorithms that have been
developed for reconstructing the density matrix directly from discrete measured
data \cite{Ar}.

To illustrate this procedure, we have simulated numerically the
reconstruction of a quantum state. 
We assumed the system is prepared initially in a
"Schr\"odinger--cat" state of the form
\begin{equation}
\label{cat}
|\Psi\rangle = \frac{1}{\sqrt{2(1+e^{-2|\alpha|^{2}} )}} (|\alpha\rangle + |-\alpha\rangle),
\end{equation}
where $|\alpha\rangle$ is a coherent state. This is a highly non--classical
state, and can be easily produced in the trapped ion system \cite{Po95}. 
In Fig.~2 we plot the real part of the reconstructed matrix elements $\rho_{n,m}$
corresponding to the initial state Eq.~(\ref{cat})
with $\alpha=1.5$. We
have taken the values of $P(x,\theta)$ for a
set of points $(x_i,\theta_j)$ and with a finite $\epsilon$.  Starting
from these data we have reconstructed the state $\rho_{n,m}$ using the
algorithm of Leonhardt {\it et al} \cite{Leb95}.  We have selected a
uniform grid of $N_x$ points corresponding to values of $x$ ranging
between $\pm 4[\hbar/(m\nu)]^{1/2}$, and a uniform grid of
$N_\theta$ points for $0\le \theta < \pi$.  Figure 2(b) corresponds
to a squeezing parameter $|\epsilon|=2$ and a grid $N_x \times N_\theta= 30 \times 30$ points.
The reconstructed state is indistinguishable from the original one. 
We have checked that even for $|\epsilon|=1$ the
obtained state is remarkably similar to the original one. We have
also tested the dependence of the reconstruction on the number of grid
points.  In Fig.~2(b) we have taken a grid of 
$N_x \times N_\theta= 30 \times 15$ points,
keeping the squeezing parameter $|\epsilon|=2$. In this case, the reconstruction
is also quite faithful. 
Reducing the number of grid points causes  
small residual background structures. On the other hand, 
reconstructing density operators that involve higher Fock states
(increasing $\alpha$) requires an increased range of $x$ values and
a larger number of grid points,
since it is necessary to resolve the oscillatory behavior of these states.
Finally, in a real experiment one cannot measure the probability distribution
$P(x_i,\theta_j)$ with arbitrary precision, due to the fact that
the number of measurements is always finite. 
We have simulated the
statistical error caused by the finite sampling number
by truncating
the values of $P(x_i,\theta_j)$ to one decimal digit. That is,
we have approximated each of the exact values by one of the following
numbers $0.0,0.1,\ldots,1.0$. The results of this simulation are
shown in Fig.~2(c).
In this case, the grid size is again $N_x \times N_\theta= 30
\times 30$ points, and $|\epsilon|=2$.  The reconstruction still
resembles the original one even in the presence of these
uncertainties.  Therefore, it would be enough to perform about 100
measurements per grid point to obtain the density matrix.  Obviously,
states with a larger phonon number will require more
measurements.

In summary, we have presented a scheme to measure the quantum state of
motion for a single ion confined in a harmonic potential (statistical 
mixtures as well as pure states). It is based on the
detection of the ground state population of the trap after a sudden change
of the trapping potential. We wish to emphasize that the effect of a sudden 
displacement of the trap center and a sudden opening of the trap can be
obtained (in principle) by a single non-instantaneous process that yields the 
same symplectic phase space transformation. Moreover, these two 
operations can be mimicked using Raman pulses with two lasers
of frequencies differing
by $\nu$ and $2\,\nu$, respectively \cite{NISTExperim,Ci93b}. Note, however,
that in this case the scheme only works in the Lamb--Dicke limit. 
Finally, it is straightforward to generalize the schemes presented
here to measure the quantum state of motion in two and three 
spatial dimensions.
This can be done by moving and opening the trapping potential 
along different directions.

We thank U.~Leonhardt, C.~Monroe,  D.~Wineland and W.~Schleich
for inspiring discussions.
J.F.P. and J.I.C. acknowledge hospitality at the University of Innsbruck.
J.F.P. is supported by a grant of the Junta de Comunidades de Castilla--La
Mancha. Part of this work has been supported by the Austrian Science
Foundation.



\begin{figure}
\caption
{
Phase space representation of the operations (phase shifting, 
displacement and squeezing) required to measure the
$Q(\alpha)$ function (a) and the quadrature distribution~(b);
(c) Level scheme and laser configuration for the detection of the 
trap ground state population. 
}
\end{figure}

\begin{figure}
\caption
{
Real part of the reconstructed density matrix elements $\rho^R_{n,m}$ 
for a ``Schr\"odinger--cat'' state with $\alpha=1.5$;
(a)~$N_x\times N_\theta=30\times 30$ and $|\epsilon|=2$ ;
(b)~$N_x\times N_\theta=30\times 15$ and $|\epsilon|=2$;
(c)~same parameters as in (a), but with $P(x,\theta)$ rounded
to one decimal digit. 
}
\end{figure}

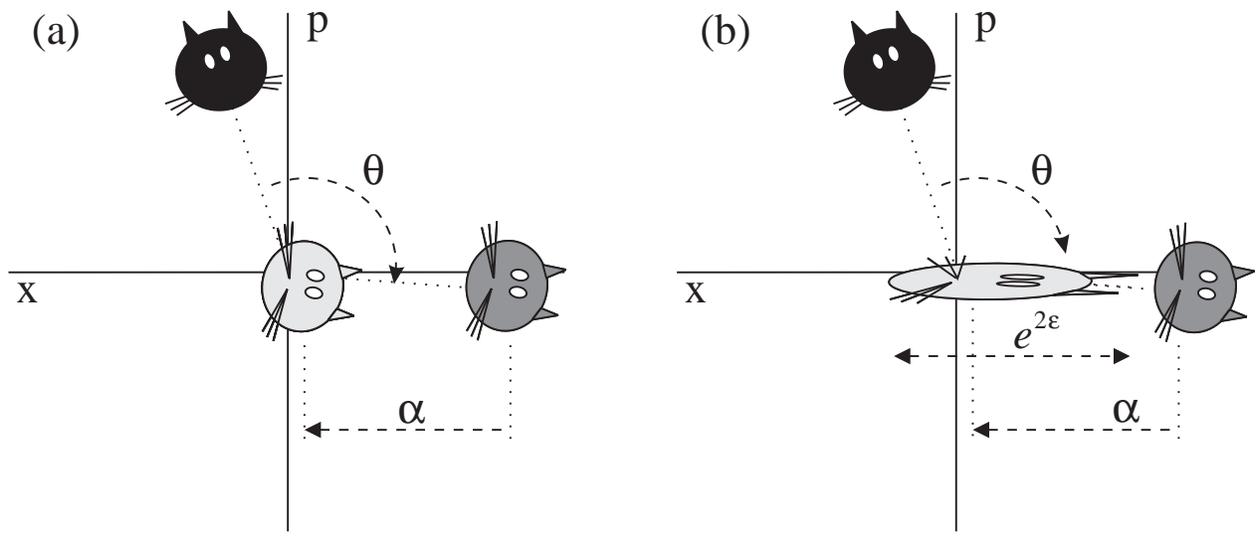
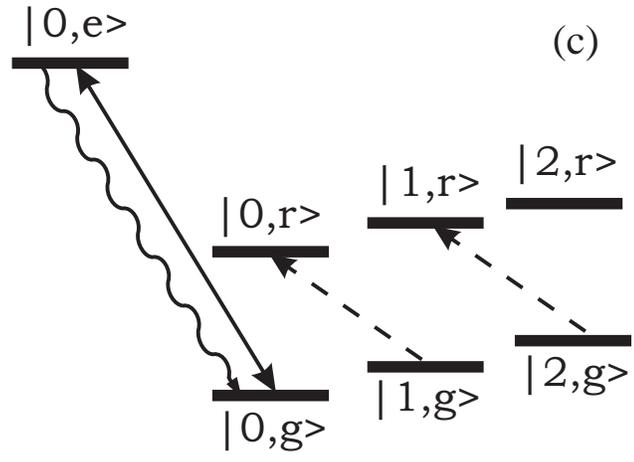
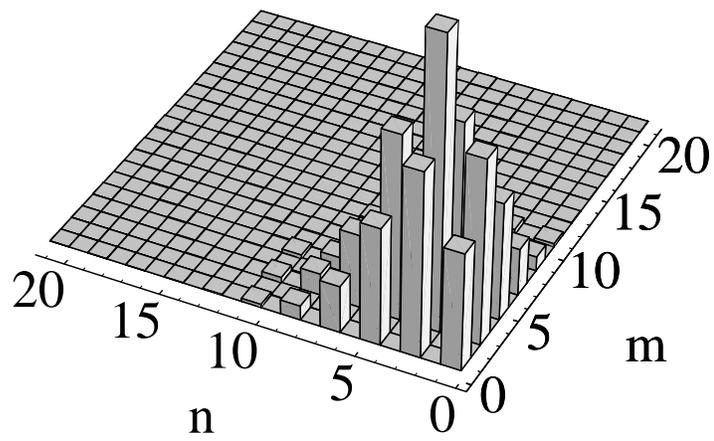
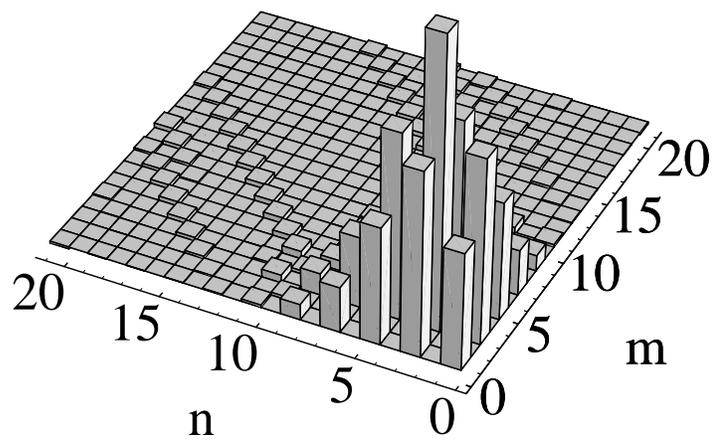
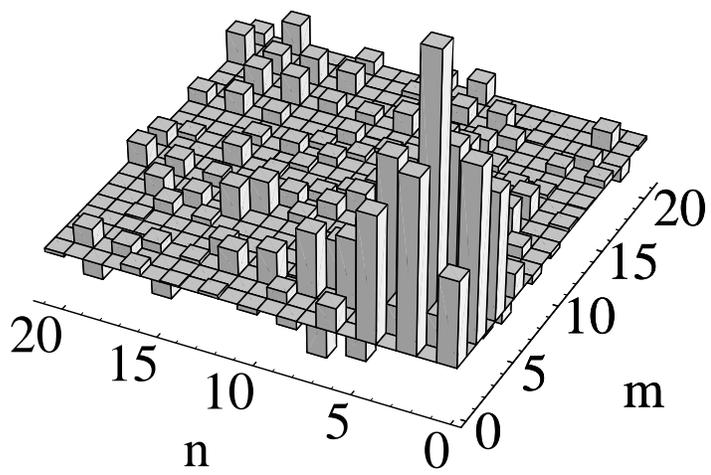
\end{document}